\documentstyle[aps,preprint,epsf,multicol]{revtex}

\def\ket#1{\left | {#1} \right >}
\def\bra#1{\left < {#1} \right |}
\def\avg#1{\left < {#1} \right >}
\def\ie{{\it i.e.}}
\def\etal{{\it et. al. }}
\def\prb{Phys. Rev. B }
\def\prl{Phys. Rev. Lett. }

\begin{document}

\draft
\title{Wave propagation through a coherently amplifying random medium}
\author{Sandeep K. Joshi\cite{jos} and  A. M. Jayannavar\cite{amj}  }
\address{ Institute of Physics, Sachivalaya Marg, Bhubaneswar 751 005, India}

\maketitle

\begin{abstract} 

We report a detailed and systematic numerical study of wave propagation
through a coherently amplifying random one-dimensional medium. The coherent 
amplification is modeled by introducing a uniform imaginary part in the
site energies of the disordered single-band tight binding Hamiltonian. 
Several distinct length scales (regimes), most of them new, are 
identified from the behavior of transmittance and reflectance as a 
function of the material parameters. We show that the transmittance is a 
non-self-averaging quantity with a well defined mean value. The 
stationary distribution of the super reflection differs qualitatively 
from the analytical results obtained within the random phase 
approximation in strong disorder and amplification regime. The study of 
the stationary distribution of the phase of the reflected wave reveals 
the reason for this discrepancy. The applicability of random phase 
approximation is discussed. We emphasize the dual role played by the 
lasing medium, as an amplifier as well as a reflector. 


\end{abstract}

\section{INTRODUCTION}

Wave propagation through a random passive medium is being studied
intensively over several decades\cite{tvr,ping}. It is now well
established that coherent interference effects, due to elastic scattering
by the static disorder, induces Anderson localization for quantum as well
as classical waves. Some physical examples are electron transport in
disordered conductors, light or electro-magnetic wave propagation in
random dielectric media, sound propagation in an inhomogeneous elastic
medium, etc. These qualitatively different types of waves in an
appropriate limit follow the same mathematical equation, namely, the
Helmholtz equation. Thus studies on different types of wave propagation
complement each other. It is basically the wave character leading to
interference and diffraction which is the common operative feature. 
Coherent multiple scattering of a wave from a fixed spatial realization of
randomly distributed scatterers generates an interference pattern in space
which is very sensitive to the actual distribution of scatterers.  Small
relative displacement of scatterers, of the order of a fraction of a
wavelength, can alter completely the interference pattern. Specifically in
the context of quantum electron transport in one dimensional random media
this will make the resistance ( or the transmittance ) a non-self
averaging quantity \cite{fuku,nkuamj} in that the resistance fluctuations over
the ensemble of macroscopically identical samples dominate the ensemble
average, \ie root mean square variation of sample to sample fluctuations
in the resistance over all the realizations of the macroscopically
identical samples exceeds the mean value by orders of magnitude no matter
how large the sample is. The inelastic scattering ( due to phonons or
other quasi-particles ) lead to loss of phase memory of the wave function. 
Thus the motion of electrons becomes phase incoherent and sample to sample
fluctuations become self-averaging in the high temperature limit leading
to a classical behavior.  In recent years the study of wave propagation
in an active random medium
\cite{expt,Nkupp,weav,abhi,pass,zhang,beenak,asen,pusti,misir,yosef,freil1,john,genak,garcia,freil2,joshi},
\ie, in the presence of absorption or amplification, is being pursued
actively. The light propagation in an amplifying (lasing) medium has its
implications for stimulated emission from random media. In a stimulated
emission photons emitted will have the same frequency, phase, direction,
and polarization. This will result in spatial and temporal coherence of
laser light propagation. 

The absorption in the medium corresponds to the actual removal of the
particle ( or energy in the case of electro-magnetic wave propagation ) by
re-combination processes. For example propagation of optical (excitons) or
magnetic excitations in solids which terminate upon reaching trapping
sites. To allow for the possibility of inelastic decay on the otherwise
coherent tunneling through potential barriers several studies invoke
absorption\cite{stone,zohta}. In the presence of inelastic scattering due
to thermal excitations, electrons are scattered out of elastic channel to
other inelastic channels. In these studies the absorption is identified as
the spectral weight lost in the inelastic channels. As an example, in the
case of one-dimensional double barrier structures the absorbed or
attenuated part is assumed to tunnel through both the left and the right
hand sides of the barriers in proportion to the transmission coefficient
of each barrier, and this is added to the coherent transmission to get the
overall transmission coefficient \cite{zohta}. In the electro-magnetic
wave propagation the bosonic nature of photons brings in both features,
namely that of amplification as well as attenuation. Photons obey Bose
statistics and their number is not conserved. Thus one can consider a
problem of wave propagation in a coherently amplifying (or absorbing)
optical medium. In the Schr\"odinger equation, to describe the absorption
or amplification, one introduces the imaginary potentials. In that case
the Hamiltonian becomes non-Hermitian and thus the particle number is not
conserved. Such Hamiltonians are widely used in Nuclear physics literature
and the corresponding imaginary potentials are called optical potentials.
The absorption or amplification for the case of light propagation is
simulated via the imaginary part of the dielectric constant with opposite
signs. It should be noted that in quenched random systems with imaginary
potentials the temporal coherence of the wave is preserved in spite of
amplification or absorption which causes a particle non-conserving
scattering process. Almost all the studies reported so far have considered
a linear amplifying or absorbing medium, irrespective of the fact that
real problem of laser oscillations and mode selection in an optically
pumped random medium requires consideration of non-linearities
\cite{Nkupp}. In all these studies the basic issue is to understand the
interplay of phase coherent multiple scattering and amplification (or
absorption). 

Several new results have been obtained from the studies of wave
propagation in an active medium. In earlier studies it has been widely
thought that the effect of absorption on classical waves is analogous to
that of inelastic scattering of electrons. Weaver\cite{weav} has shown
that absorption does not provide a cut-off length scale ( similar to an
inelastic scattering length ) for the renormalization of wave transport in
the random medium. In other words, the absorption does not re-establish
the diffusive behavior of the wave propagation by destroying the
localization of eigenfunctions. The transport seems to remain
non-diffusive even in the presence of absorption. This fact will have an
important bearing on the physics at the mobility edge in higher
dimensional (3-D) systems. In a related development it has been shown that
absorption along with enhanced reflection induces coherence in quantum
systems\cite{amj1}. In a scattering problem, the particle experiences a
mismatch from the real valued potential to the imaginary valued potential
at the interface between the free region and the absorbing (or amplifying)
medium, and hence it tries to avoid this region by enhanced back
reflection. Thus a dual role is played by imaginary potentials as an
absorber (or amplifier) and as a reflector. This point has been emphasized
in earlier treatments\cite{abhi,amj1,ruby}. One can readily show that,
when the strength of the imaginary potential is increased beyond certain
limit, both absorber and amplifying scatterer act as a reflector. Thus the
reflection coefficient exhibits non-monotonic behavior as a function of
the absorption (amplification) strength. Using the duality relations it
has been shown that amplification suppresses the transmittance in the
large length (L) limit just as much as absorption does \cite{pass}. This
is somewhat contrary to the expectation. One would have expected that as a
wave passes through a disordered amplifying medium it undergoes coherent
multiple scattering and hence gets amplified before it escapes from the
system. It turns out that coherent amplification in turn induces
localization by enhancement of the coherent backscattering involving
longer return paths, thereby cutting off transmission. Experimentally this
reflects in a narrowing of the backscattering cone in random amplifying
medium \cite{expt}. It has been noticed \cite{zhang} that there exists a
crossover length scale $L_c$ below which the amplification enhances the
transmission and above which the amplification reduces the transmission
which, in fact, vanishes exponentially in the $L \rightarrow \infty$
limit. In contrast, super-radiant reflectance saturates to a finite value
in the large length limit. Moreover, absorption and amplification of same
strength (\ie, differing only in the sign of the imaginary part) will
induce same localization length \cite{beenak}.  Even for an ordered
periodic system in the presence of coherent amplification, the
transmittance always decreases in the asymptotic length limit \cite{asen}.
This follows from the fact that the amplifier also acts as a
back-scatterer (or reflector) as mentioned above. To obtain enhanced
coherent transmittance, the synergy between wave confinement due to Anderson
localization and coherent amplification by active medium is not necessary.
By a proper choice of a length of an ordered amplifying medium one can
achieve large transmittance. However, for a finite sample of length $L$ to
obtain enhanced reflection the synergy between disorder and amplification
plays a major role.

In an amplifying medium even though the transmittance ($t$) decreases
exponentially with the length $L$ in the large $L$ limit, the average
$\avg{t}$ is shown\cite{pusti} to be infinite due to the less probable
resonant realizations corresponding to the non Gaussian tail of the
distribution of $ln~t$. This result is based on the analysis using random
phase approximation (RPA). Using duality argument Paasschens \etal show
that non-Gaussian tails in the distribution of $lnt$ contain negligible
weight\cite{pass}. Thus one might expect finite value for $\avg{t}$ in the
asymptotic limit. It should be noted that even in the ordered periodic
system all the states are resonant states and still the transmittance
decreases exponentially for all the states in the large $L$ limit. The
above simple case may indicate that in the asymptotic limit $\avg{t}$ is
indeed finite. One of our objectives in this paper is to study
(numerically) the behavior of the transmission probability as a function
of length in the presence of coherent amplification.  We show that the
transmission coefficient is a non-self-averaging quantity. In the large
length limit we do not find any resonant realization, which can give an
enhanced transmission. We also study the behavior the of cross-over length
$L_c$ as a function of disorder and amplification strength. As mentioned
earlier, upto the cross-over length $L_c$ transmittance increases and
after $L_c$ it falls exponentially. We have studied the logarithm of the
transmittance which will have a maximum value $\avg{lnt}_{max}$ at $L_c$. 
We have analyzed the behavior of $\avg{lnt}_{max}$ as a function of
disorder and amplification strength. We would like to emphasize that in
the lasing medium the presence of disorder suppresses the average
transmittance at all length scales in comparison with the ordered media
having the same strength of amplification. For a given strength of
amplification there exists a critical strength of disorder below which the
average transmittance is always less than unity at all length scales and
decreases monotonically. In this regime $L_c$ and $\avg{lnt}_{max}$ loose
their physical significance. Yet in this regime we show that there exists
a cross-over length scale $\xi_c$ which diverges as the amplification
strength is reduced to zero for a given strength of the disorder. In the
case of super reflection in the presence of disorder we show that there
exists a cross-over length $L_1$ below which the averaged logarithm of
reflectance, $\avg{lnr}$, is always less than $lnr$ for the periodic
($W=0$) lasing ($\eta~\neq~0$) system. $L_1$ depends on disorder and
amplification strength. Below $L_1$, $\avg{lnr}$ is always larger than
that for the ordered lasing medium. However, there is another disorder
dependent length scale $L_2 ~<~ L_1$. For a system of size less than
$L_1$ disorder enhances the reflection whereas for sizes between $L_2$ and
$L_1$ disorder suppresses the reflection. 

In the work by Pradhan and Kumar\cite{Nkupp}, the analytical expression for the stationary
distribution $P_s(r)$ of a coherently backscattered reflection coefficient $(r)$
is obtained in the presence of both absorption and amplification using the method
of invariant imbedding\cite{rrbd}. In the presence of a spatially uniform amplification 
in a random medium and with the help of random phase approximation, the 
expression for $P_s(r)$ is given by

\begin{eqnarray}
P_s(r) & = & \frac{|D|exp(-\frac{|D|}{r-1})}{(r-1)^2}~~~for~~ r \geq 1 \label{prad} \\
       & = &  0~~~~~~~~~~~~~~~~~                     ~~~for~~ r < 1 \nonumber
\end{eqnarray}
where $D$ is proportional to $\eta / W$, $\eta$ and $W$ being the 
strength of amplifying potential and disorder respectively. One can readily
notice from Eqn. (\ref{prad}) that $P_s(r)$ does not tend to $\delta (r-1)$ in
the large $\eta$ limit. In this limit, as mentioned earlier, an amplifying 
scatterer acts as a reflector. Instead, Eqn. (\ref{prad}) indicates that as $D$
increases the distribution becomes broad and the most probable value of the reflection
coefficient shifts to higher values. Since the above expression for $P_s(r)$
is obtained within the RPA, its validity is
limited for small disorder and amplification strength. Even in the absence
of amplification it is well known that RPA fails in the large disorder limit\cite{allan,amj2,amj3}.
In the presence of small disorder it is shown that absorption suppresses
phase fluctuations making the regime of validity of RPA still smaller in the parameter
space of disorder and absorption\cite{abhi,freil2}. We show that this is true also in the presence of
spatially uniform amplification. From Eqn. (\ref{prad}) it follows that the average reflection
coefficient $\avg{r}$ is infinite.

With the help of transfer matrix approach we have studied the distribution and 
statistics of the transmittance $t$ from which the non-self-averaging nature of
$t$ follows. We then study the behavior of
$L_c$, $L_1$, $L_2$ and $\avg{lnt}_{max}$ on the material parameters.
The probability distribution of the reflection coefficient $P(r,L)$ tends
to a stationary distribution $P_s(r)$ in the large $L$ limit. For a small
disorder $W$ and small amplification $\eta$, $P_s(r)$ is qualitatively in
conformity with Eqn. (\ref{prad}). As we increase $\eta$, a double peak structure
appears in $P_s(r)$ and as we increase $\eta$ further, $P_s(r)$ tends towards
$\delta (r-1)$. The average of $lnr$ obtained from $P_s(r)$ exhibits 
maxima as a function of $\eta$.
We also show that amplification suppresses the phase fluctuation of the complex
reflection amplitude. In the next section we define our model Hamiltonian and 
transfer matrix approach. Later sections are devoted to results and conclusions.

\section{NUMERICAL PROCEDURE}

We consider a quasi-particle moving on a lattice. The appropriate Hamiltonian in
a tight-binding one-band model can be written as 

\begin{equation}
H~=~\sum \epsilon_n^\prime \ket{n}\bra{n} + V(\ket{n}\bra{n+1}~+~\ket{n}\bra{n-1} ).
\end{equation}

$V$ is the off-diagonal matrix element connecting nearest neighbors separated
by a lattice spacing $a$ (taken to be unity throughout) and $\ket{n}$ is the
non-degenerate Wannier orbital associated with site $n$, where $\epsilon_n^\prime
=\epsilon_n-i\eta$ is the site energy. The real part of the site energy
$\epsilon_n$ being random represents static disorder and $\epsilon_n$ at
different sites are assumed to be uncorrelated random variables distributed
uniformly ($P(\epsilon_n)=1/W$) over the range $-W/2$ to $W/2$. Here $W$ can
be interpreted as the strength of the disorder. We have taken imaginary part of the
site energy $\eta$ to be spatially uniform and depending on whether the medium
is amplifying or absorbing, it is set to positive or negative values. Since all
the relevant energies can be scaled by $V$, we can set $V$ to unity. The lasing
medium consisting of $N$ sites ($n=1$ to $N$) is embedded in a perfect infinite
lattice with all site energies taken to be zero.

To calculate the transmission and reflection coefficients we use the well known
transfer-matrix method \cite{chao}. Here we describe the method very briefly.
Let the sample be placed between two semi-infinite perfect leads. With the
wave-function $\psi$ inside the sample expressed as a linear combination of the 
Wannier orbitals $\left . | n \right >$ with coefficients $c_n$ the 
Schr\"odinger equation $H\psi~=~i\dot \psi$ leads to
\begin{equation}
\left [
\begin{array}{c} c_{n+1}\\ c_n \end{array}\right ]~=~ \left [
\begin{array}{cc} \frac{(E-\epsilon_n^\prime)}{V} & -1 \\ 1 & 0 \end{array} 
\right ] \left [ \begin{array}{c} c_n \\ c_{n-1} \end{array} \right ]
\end{equation}
where $E$ is the energy of the incident particle. Thus to obtain all the coefficients
$c_n$, for $n=1$ to $N$, we just have to evaluate the product of $N$ $2\times2$ 
matrices $T_i$ of the type shown above. If a plane wave $e^{ikn}$ is sent 
through the perfect lead from one side then the solutions on the two sides
of the sample are related by a product matrix $M$ \ie,
$$ M~=~\omega S^{-1}\prod^{N}_{i=1} T_i S,$$
where
$$\omega ~=~ \left [ 
\begin{array}{cc} e^{ik(N+1)} & 0 \\ 0 & e^{-ik(N+1)} \end{array} \right ],~~~~
S~=~\left [ \begin{array}{cc} e^{-ik} & e^{ik} \\ 1 & 1 \end{array} \right ]
$$
The transmission $(t)$ and reflection $(r)$ coefficients in terms of the 
matrix elements of $M$ are 
$$ t ~ = ~ \frac{detM}{|M_{11}|^2},~~~~r~=~\frac{|M_{12}|^2}{|M_{11}|^2}.$$
Since the Hamiltonian is non-hermitian we have $t+r\neq 1$.

\section{RESULTS AND DISCUSSION}

In our studies we have set the energy of the incident particle 
at $E=0$, \ie, at a midband energy. Any other value for the incident
energy does not affect the physics of the problem. In calculating average
values in all cases we have taken 10,000 realizations of random site energies
($\epsilon_n$). The strength of the disorder and the amplification are scaled
with respect to $V$, \ie, $W~(\equiv W/V)$ and $\eta~(\equiv \eta/V)$.
The length $L$ denotes the dimensionless length in the unit of
lattice spacing $a$.

\begin{figure}
\protect\centerline{\epsfxsize=6in \epsfbox{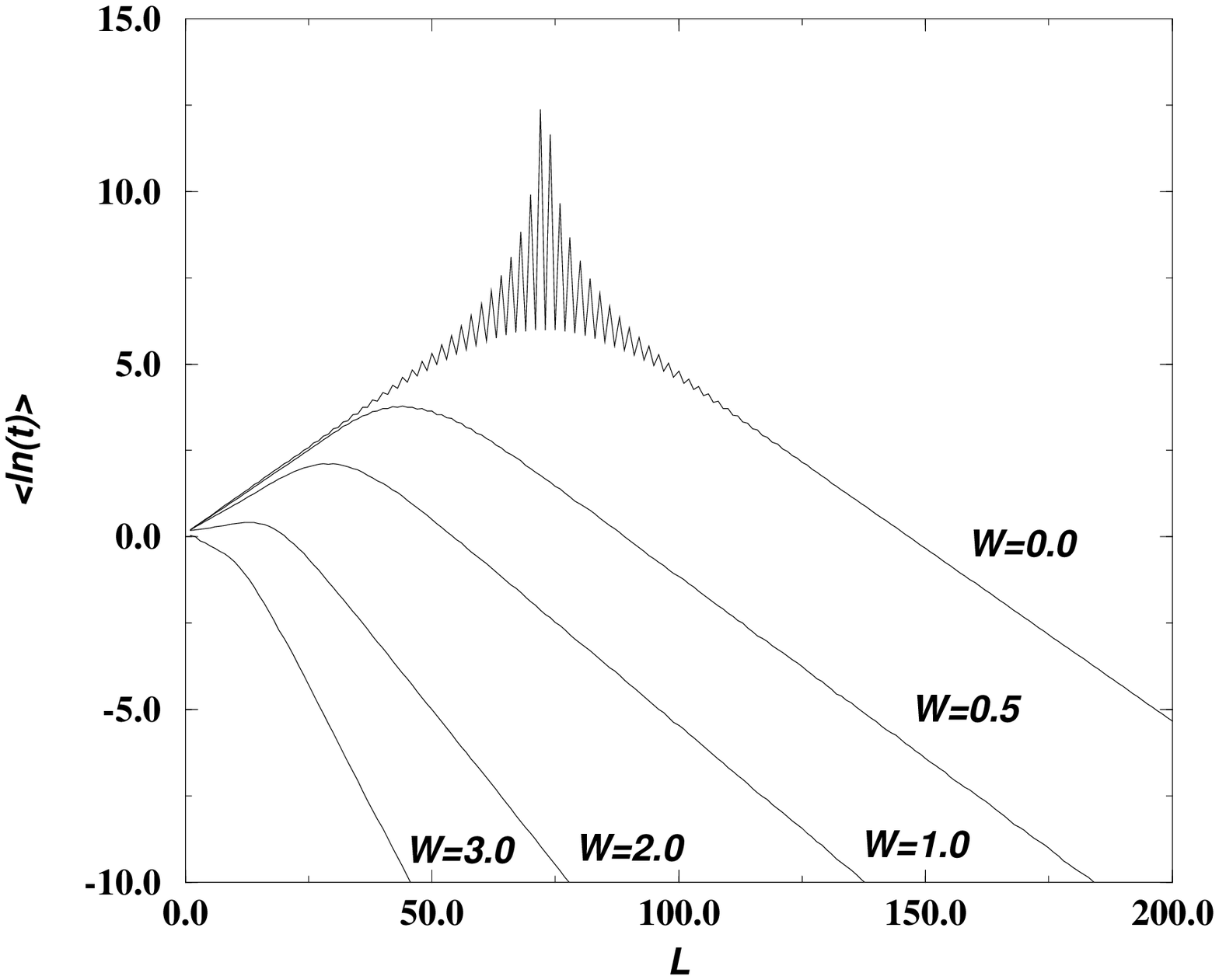}}
\caption{Average of logarithm of transmission coefficient $t$ versus
length $L$ for $\eta=0.1$ and different values of $W$.}
\label{lntW} 
\end{figure}

Depending on the parameters $\eta$, $W$ and $L$ the transmission coefficient
can be very large (of the order of $10^{12}$ or more). Hence, we first
consider behavior of $\avg{lnt}$ instead of $\avg{t}$. The angular
brackets denote the ensemble average. In Fig. \ref{lntW} we have 
plotted $\avg{lnt}$ as a function of the length $L$ for a fixed value
of amplification strength $\eta=0.1$ and for various values of the disorder
strength $W$ as indicated in the figure. In the absence of disorder
($W=0$) as one varies the length, initially the transmission increases
to a very large value ($t \approx 10^{12}$) through large oscillations and 
after exhibiting a maximum at the length $L_c$, and again through
oscillations, it eventually decays exponentially to zero as $L\rightarrow
\infty$. We denote the maximum in $\avg{lnt}$ at $L=L_c$ 
as $\avg{lnt}_{max}$, $L_c$ being the cross-over length. In the presence of 
disorder one
readily notices that $\avg{lnt}$ is suppressed at all
lengths as compared to an ordered amplifying medium of same $\eta$.
Both $L_c$ and $\avg{lnt}_{max}$ decrease as functions of the disorder 
strength. When the disorder strength is large (see Fig. \ref{lntW}
for $W=3.0$) the average transmittance becomes less than unity and
decreases monotonically as a function of $L$. In this regime both
$L_c$ and $\avg{lnt}_{max}$ lose their physical significance. We will show
later that even in this regime one can still define a new cross-over
length scale, say $\xi_c$. The existence of $L_c$ can be 
attributed to the synergetic effect between the amplification and the
localization. Eventually the strong back scattering arising 
due to both serial one dimensional disordered potential and amplification
leads to an exponential decay of the transmittance. From the graphs
of $\avg{lnt}$ versus $L$, one can find the corresponding localization
length. We denote the localization length by $l_a$ for an ordered medium ($W=0$)
in the presence of uniform amplification. The localization length \cite{econ} for
a disordered passive medium ($\eta=0$) is given by elastic back scattering
length $l=48V^2/W^2$ at the center of the band ($E=0$). We have verified
that the localization length in the presence of both disorder and
amplification,\cite{zhang} $\xi$ is related to $l$ and $l_a$ (for $\eta/V < 1$
and $W/V < 1$) as $\xi=ll_a/(l+l_a)$. We have also verified that
$\xi(\eta)=\xi(-\eta)$ as shown by Paasschens \etal using duality 
argument\cite{pass}.

\begin{figure}
\protect\centerline{\epsfxsize=6in \epsfbox{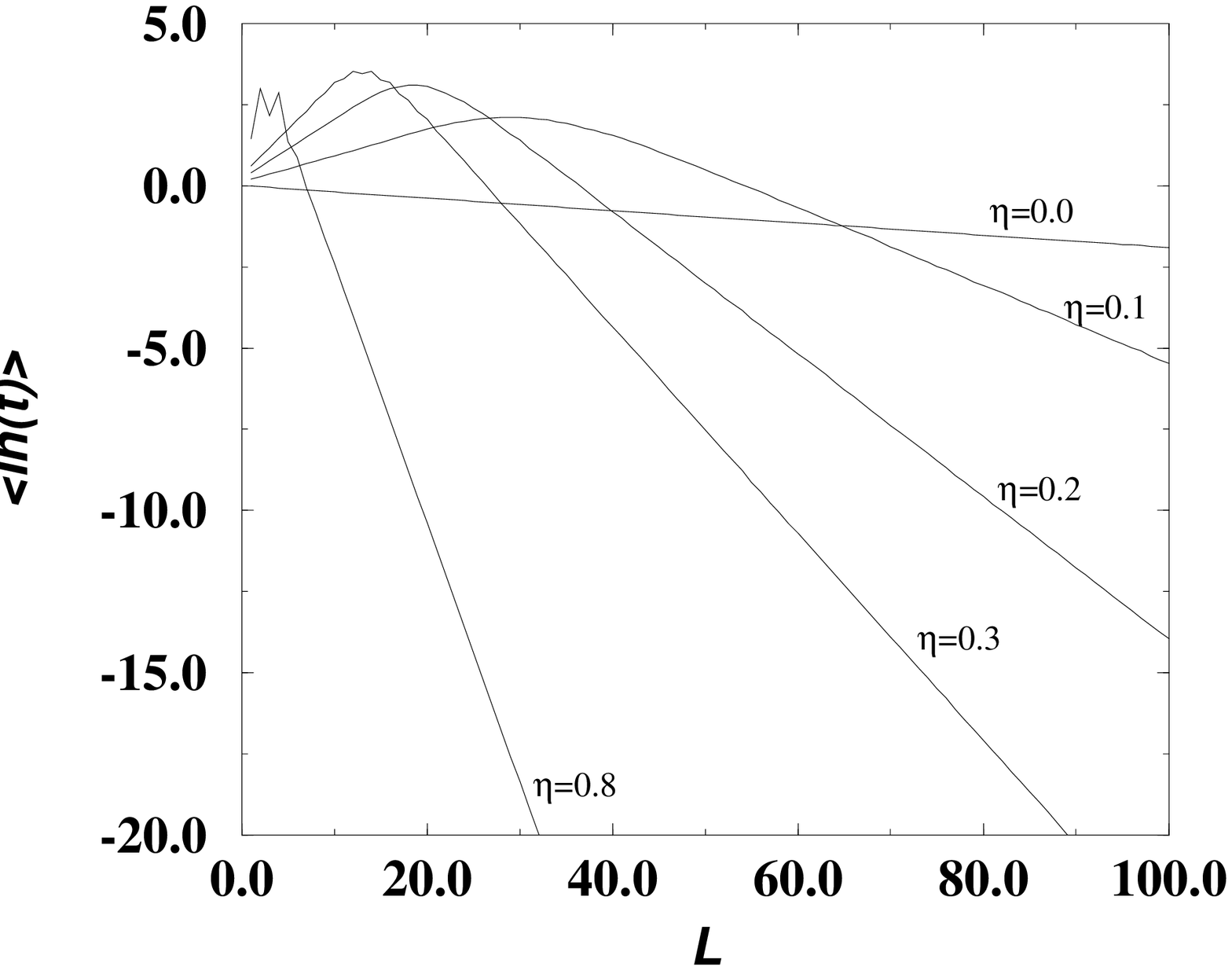}}
\caption{Average of logarithm of transmission coefficient $t$ versus
length $L$ for $W=1.0$ and different values of $\eta$.}  
\label{lntE}
\end{figure}

Fig. \ref{lntE} illustrates the behavior of $\avg{lnt}$ as a
function of $L$ for a fixed value of disorder $W=1.0$ and for various values
of the amplification strength $\eta$ as indicated in the figure. One finds that
the cross-over length $L_c$ is a monotonically decreasing function of $\eta$.
However, $\avg{lnt}_{max}$ initially increases with $\eta$ and after exhibiting 
maxima, $\avg{lnt}_{max}$ decreases with further increase in $\eta$. 

\begin{figure}
\protect\centerline{\epsfxsize=6in \epsfbox{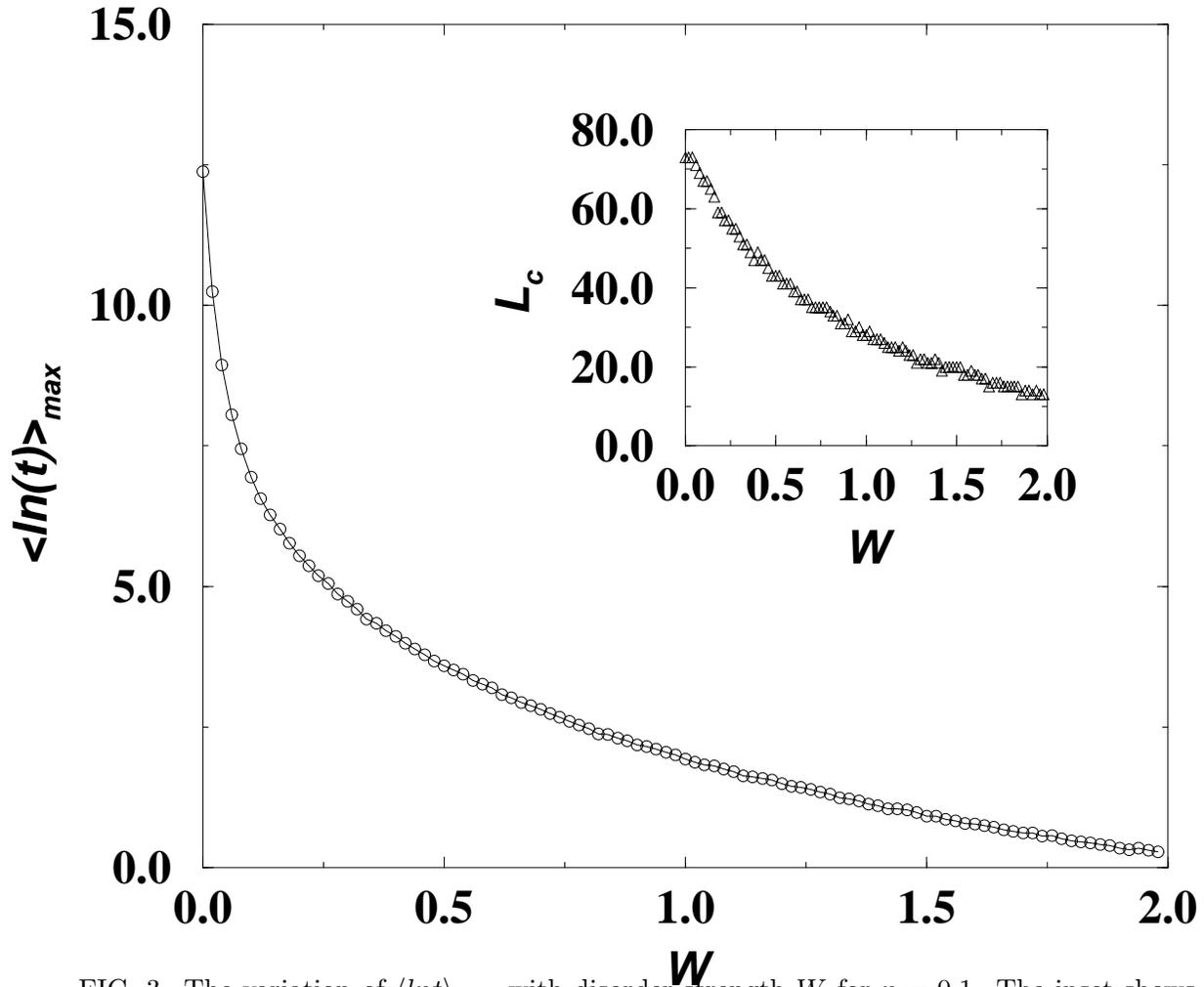}}
\caption{The variation of $\avg{lnt}_{max}$ with disorder strength $W$ for 
$\eta=0.1$. The inset shows the variation of $L_c$ with $W$ for $\eta=0.1$.}
\label{MaxTvsW}
\end{figure}

We will now study the behavior of $L_c$ and $\avg{lnt}_{max}$ as functions of $W$
and $\eta$ in the parameter space where $L_c$ and $\avg{lnt}_{max}$ are well-defined. 
In Fig. \ref{MaxTvsW} we have plotted $\avg{lnt}_{max}$ versus $W$ for a fixed
$\eta =0.1$. In the inset of Fig. \ref{MaxTvsW} is shown the dependence of
$L_c$ on $W$. It is clear that both $\avg{lnt}_{max}$ and $L_c$ monotonically
decrease with $W$. The cross-over length $L_c$ does not follow the 
$1/W$ prediction \cite{zhang}. This comes out by curve fitting our numerical 
data in the full parameter range. The prediction that $L_c \sim 1/W$ has the 
shortcoming that in the limit $W\rightarrow 0$ it tends to infinity, but we
know that for a perfect ordered lasing medium ($W=0$) $L_c$ is indeed finite.
The validity of $L_c \sim 1/W$ in the intermediate regime is not ruled out.

\begin{figure}
\protect\centerline{\epsfxsize=6in \epsfbox{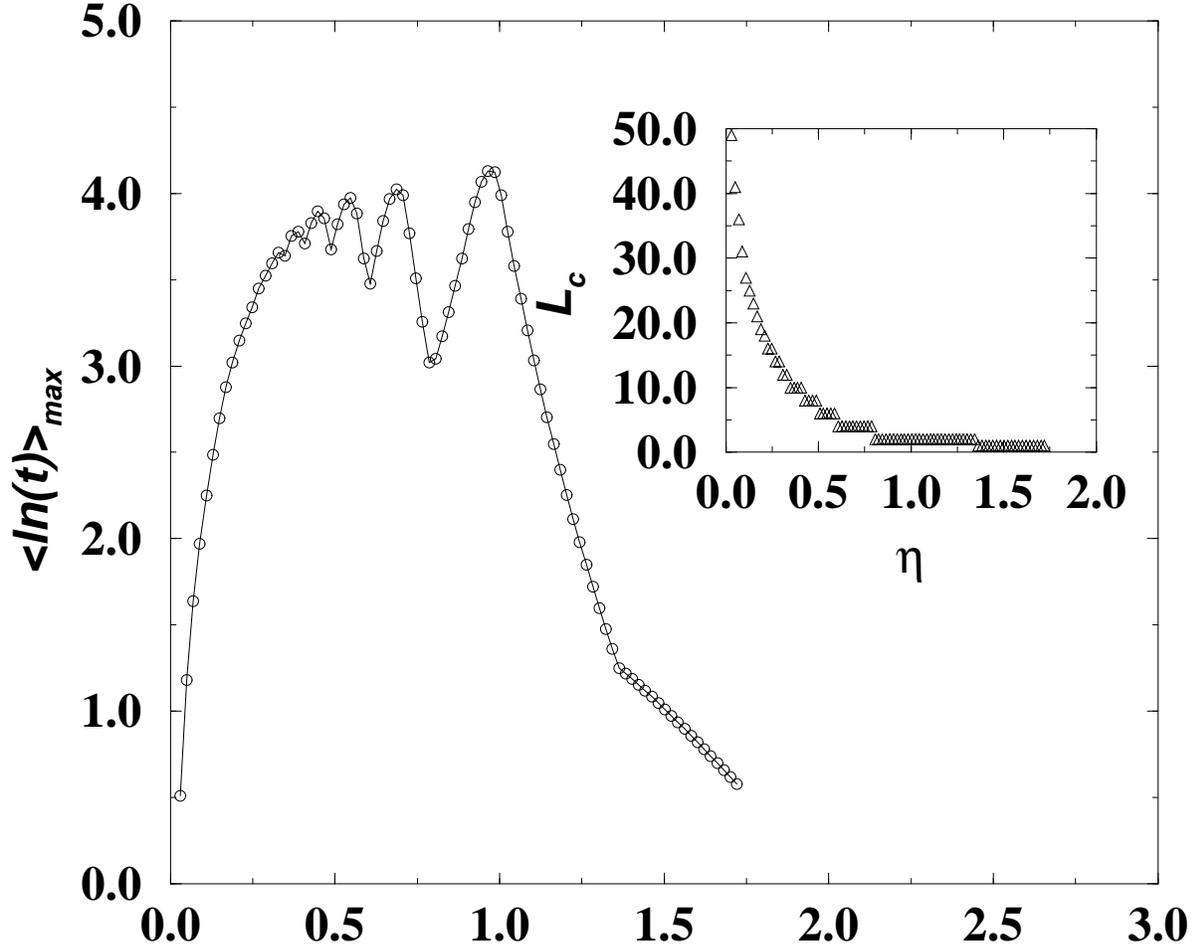}}
\caption{The variation of $\avg{lnt}_{max}$ with amplification strength $\eta$ 
for $W=1.0$. Inset shows the variation of $L_c$ with $\eta$ for $W=1.0$.}
\label{mteta}
\end{figure}

In Fig. \ref{mteta} we have plotted $\avg{lnt}_{max}$ against $\eta$ for a 
fixed value of $W=1.0$ and the inset shows variation of $L_c$ with $\eta$
for $W=1.0$. Initially $\avg{lnt}_{max}$ increases with $\eta$ and after 
exhibiting a maxima it decays to zero for large $\eta$. This arises from
the fact that the lasing medium acts as a reflector for large $\eta$ as
discussed in the introduction. Near the maximum, in a finite regime of $\eta$,
$\avg{lnt}_{max}$ exhibits several oscillations. In this region sample to 
sample fluctuations of $lnt$ are very large. Thus average over $10,000$ 
realizations may not represent the true ensemble averaged quantity. From
the curve fitting of our numerical data for $L_c$, we find that $L_c$ does not follow 
a power law, ($1/\sqrt{\eta}$), in the full parameter regime \cite{zhang}.

\begin{figure}
\protect\centerline{\epsfxsize=6in \epsfbox{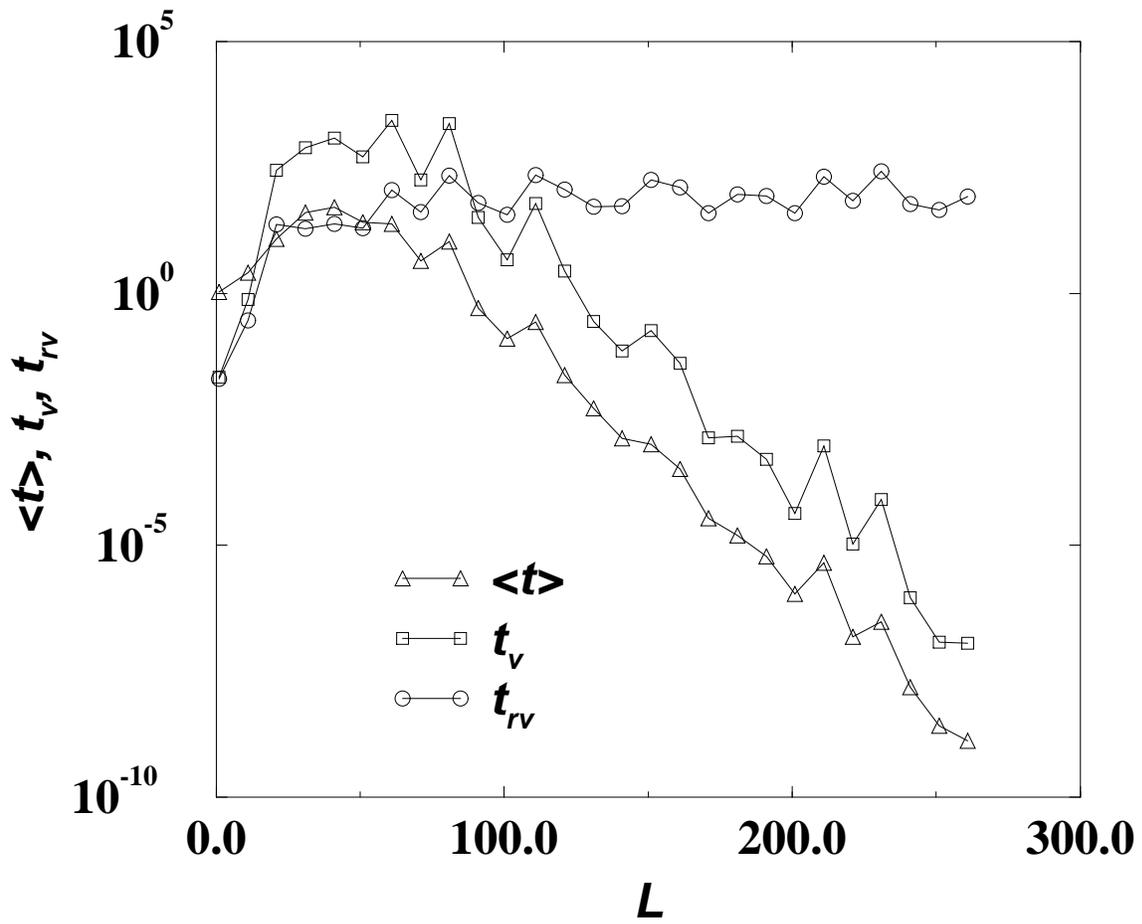}}
\caption{The plot of $\avg{t}$, root-mean-squared variance ($t_v$) and root-mean-squared
relative variance ($t_{rv}$) as a function of length $L$ for $\eta=0.1$ and $W=1.0$.}
\label{varT}
\end{figure}

\begin{figure}
\protect\centerline{\epsfxsize=6in \epsfbox{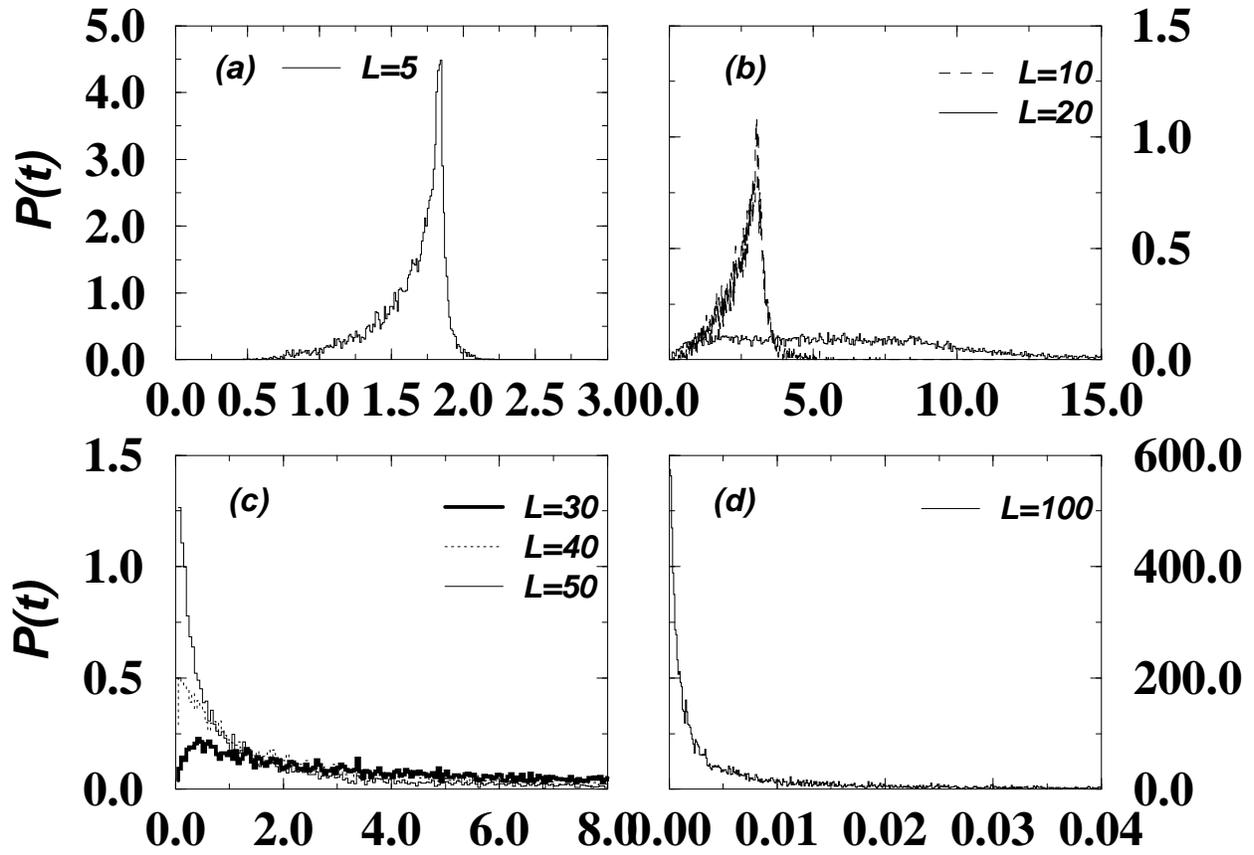}}
\caption{Distribution of transmission coefficient $t$ for $W=1.0$ and $\eta=0.1$
at different sample lengths, as indicated in the figure.}
\label{Tdist}
\end{figure}

To study the nature of fluctuations in the transmission coefficient, in Fig. 
\ref{varT} we have plotted, on log-scale, $\avg{t}$, root-mean-squared variance $t_v =
\sqrt{\avg{t^2}-\avg{t}^2}$ and root-mean-squared relative variance (or
fluctuation) $t_{rv}=\sqrt{\avg{t^2}-\avg{t}^2}/\avg{t}$ as a function of
$L$ for $\eta=0.1$ and $W=1.0$. For these parameters $l \approx 48$,$l_a = 10$,
$\xi \approx 8$ and $L_c \approx 30$. We notice that both $\avg{t}$ and $t_v$
exhibit maxima and decrease as we increase the length further. Except in the small length
limit, variance is larger than the mean value. The relative variance is larger than 
one for $L>10$ and remains large even in the large length limit. The $t_{rv}$ fluctuates
between values $50$ to $300$ in the large length ($L>10$) regime, indicating
clearly the non-self-averaging nature of the transmittance. This implies that
the transmission over the ensemble of macroscopically identical samples dominates
the ensemble average. The transmissions across the sample is very sensitive 
to the spatial realizations of impurity configurations. Because $\avg{t}$ being
non-self-averaging, it does not represent a well defined physical quantity. In
such a situation one has to consider the full probability distribution $P(t)$
of $t$ to describe the system behavior. In Fig. \ref{Tdist} we have plotted $P(t)$
as a function of $t$ for various values of $L$ as indicated in the figures.
We have chosen $W=1.0$ and $\eta=0.1$. We see that for $L<L_c(\approx 30)$
$P(t)$ is a peaked distribution with a negligible weightage at large
$t$. As we increase $L$ further the distribution broadens and the peak shifts 
to higher values of $t$ with the emergence of a tail. For larger value $L>L_c$ 
the peak again becomes sharper and starts shifting towards lower value of 
$t$ with a small weightage at tails. For further increase in $L$ a sharp 
peak appears around $t=0$ with a negligible weightage in the tail of the 
distribution.

\begin{figure}
\protect\centerline{\epsfxsize=6in \epsfbox{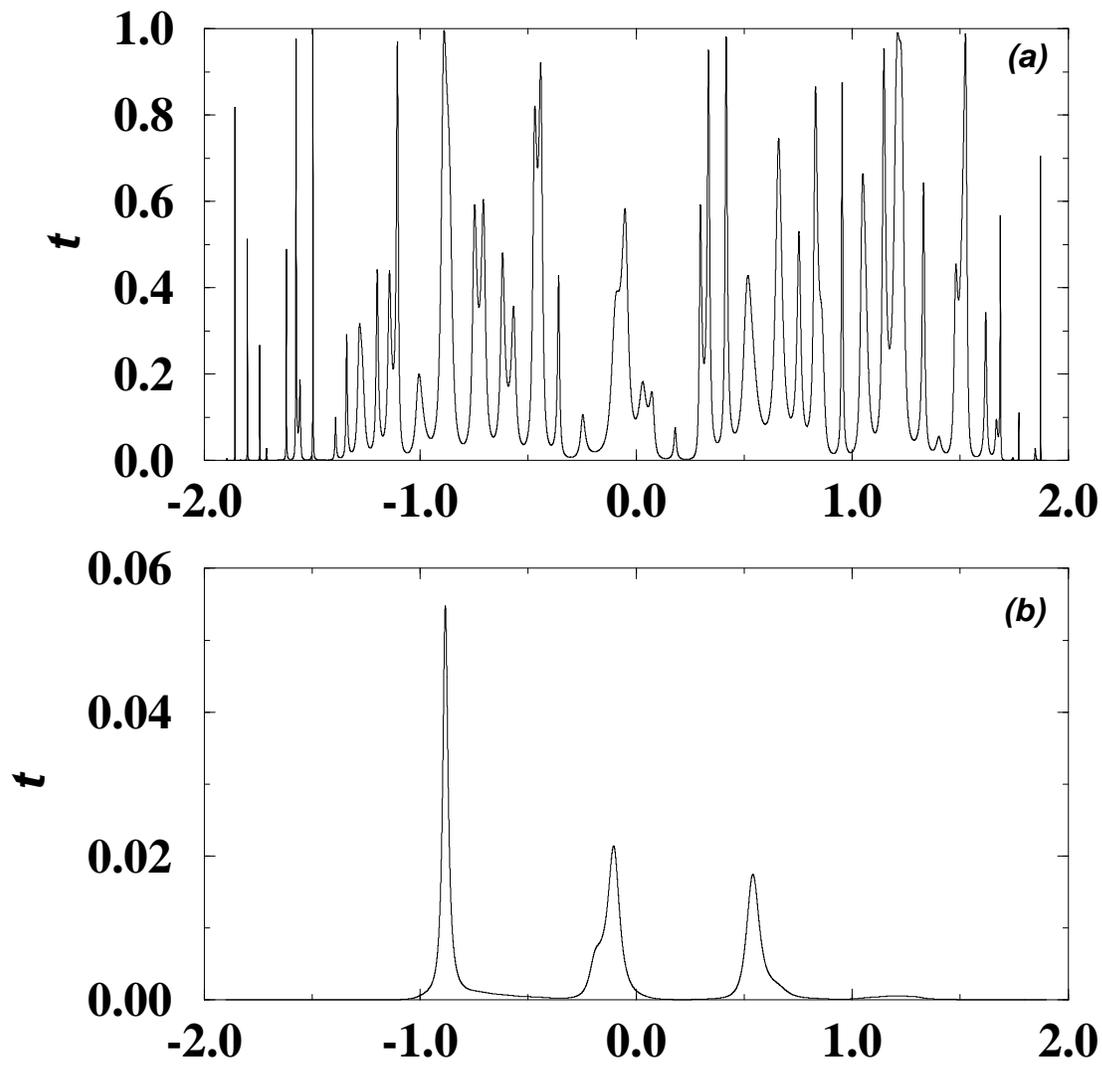}}
\caption{Transmittance $t$ as function of incident energy $E$ for $W=1.0$,
$L=100$ and (a) $\eta=0$ and (b) $\eta=0.1$}
\label{TvsE}
\end{figure}

We would now like to understand whether there exist any resonant realizations in 
the large length limit for which the transmittance is very large. This study calls
for sample to sample fluctuations. It is well known  from the studies in passive
random media that the ensemble fluctuation and the fluctuations for a given sample
as a function of chemical potential or energy are expected to be related by some 
sort of ergodicity \cite{tvr,ergod}, \ie, the measured fluctuations as a function of
the control parameter are identical to the fluctuations observable by changing the
impurity configurations. In Fig. \ref{TvsE}(a) we have plotted $t$ versus incident 
energy $E$ (within the band from $-2$ to $+2$) for a given realization of random 
potential with $\eta=0$ and $L=100$. The Fig. \ref{TvsE}(b) shows the behavior of $t$ versus
$E$ for the same realization in the presence of amplification $\eta=0.1$ and $L=100$. 
From the Fig. \ref{TvsE}(a) we observe that at several
values of energy the transmittance exhibits the resonant behavior in that $t=1$. 
These resonant states make the average of Landauer four probe conductance
( $G=(e^2/\pi\hbar)t/(1-t)$ ) infinite \cite{rolf,melnikov}. From Fig. \ref{TvsE}(b) we notice
that in the presence of amplification, transmittance at almost resonant realizations is 
negligibly small. Few peaks appear in the transmittance whose origin lies in the 
combined effect of disorder and amplification. However, we notice that the transmittance 
at these peaks is much smaller, where as one would 
have naively expected the transmittance to be much much larger than unity in the amplifying
medium. We have studied several realizations and found that none of them shows any 
resonant behavior where one can observe the large transmittance. The peak value of 
observed transmittance is of the order of unity or less. This study clearly indicates 
that $\avg{t}$ is indeed finite contrary to the earlier predictions based on RPA \cite{pusti}.

\begin{figure}
\protect\centerline{\epsfxsize=6in \epsfbox{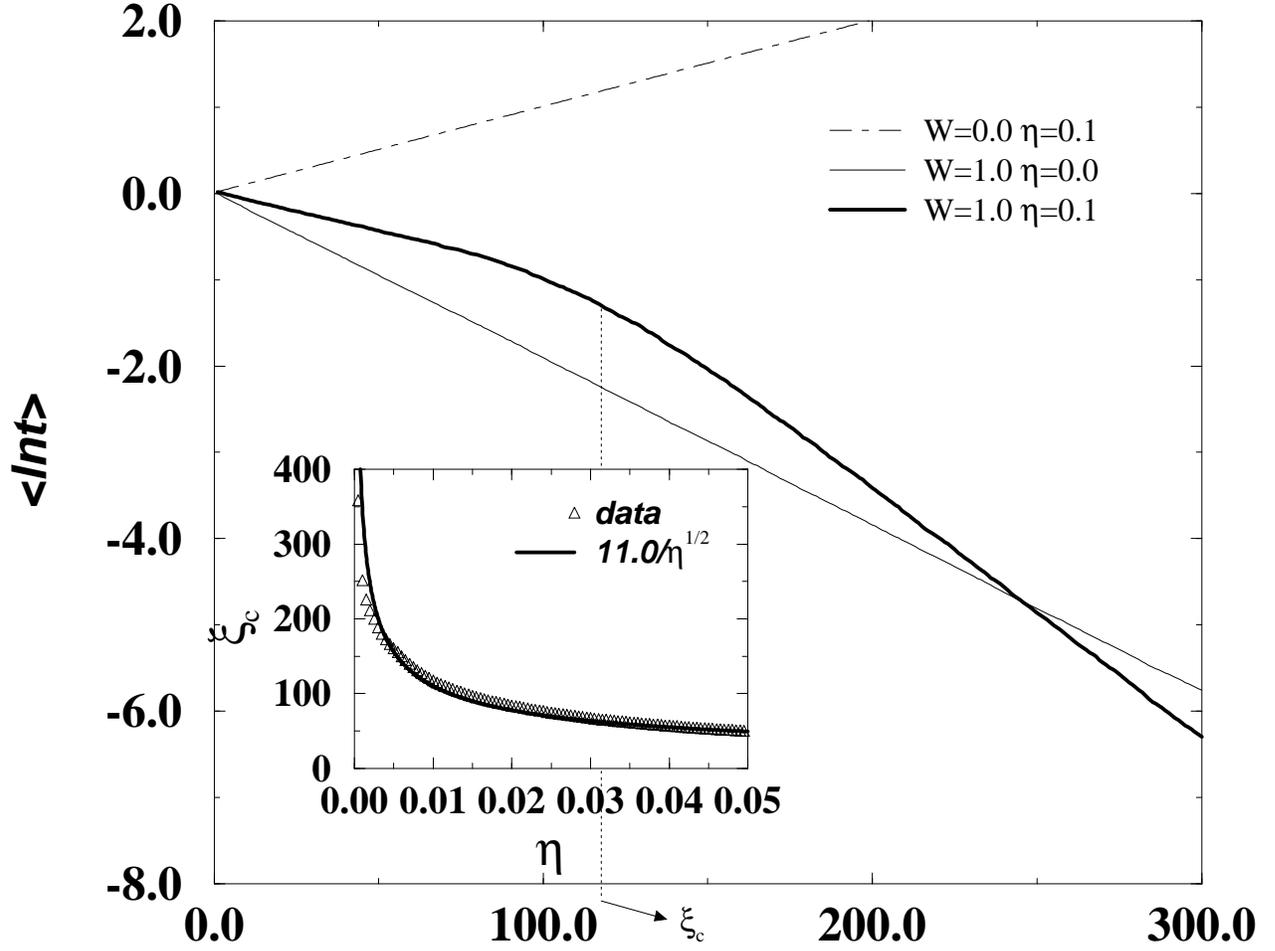}}
\caption{Variation of $\avg{lnt}$ with $L$. The new length scale $\xi_c$ which
arises for $\eta \ll 1.0$ is shown by a vertical dotted line. The inset shows
the variation of $\xi_c$ with $\eta$ for $W=1.0$. The numerical fit shown by
the thick line indicates that $\xi_c$ scales as $\eta^{-1/2}$ in this regime.}
\label{ltl}
\end{figure}

So far our study was restricted to the parameter space of $W$ and $\eta$ for
which $L_c$ and hence $\avg{lnt}_{max}$ exist. In Fig. \ref{ltl} we have plotted
$\avg{lnt}$ against $L$ for ordered lasing medium ($W=0$,$\eta=0.01$), disordered
passive medium ($W=1.0$,$\eta=0$) and disordered active medium ($W=1.0$,$\eta=0.01$).
The present study is restricted to the parameter space of $\eta$ and $W$ such that
$\eta \ll 1.0$ and $W \geq 1.0$. We notice that for an ordered lasing medium, the
transmittance is larger than one.
We have taken our range of $L$ upto $300$. Needless to say, in the asymptotic
regime, for an ordered lasing medium, the transmittance tends to zero exponentially.
For a disordered active medium ($W=1.0$,$\eta=0.01$), we notice that the transmittance
is always less than one and monotonically decreasing. Initially, upto certain
length, the average transmittance is, however, larger than that in the disordered passive
medium ($W=1.0$,$\eta=0$). This arises due to the combination of lasing with disorder.
In the asymptotic regime transmittance of a lasing random medium falls below that in the
passive medium with same disorder strength. This follows from the enhanced localization
effect due to the presence of both disorder and amplification together, \ie, $\xi < l$.
It is clear from the figure that $\avg{lnt}$ does
not exhibit any maxima and hence the question of $L_c$ or $\avg{lnt}_{max}$ does
not arise. We notice, however, from the figure that for random active medium initially
$\avg{lnt}$ decreases with a well defined slope and in the large length limit
$\avg{lnt}$ decreases with a different slope (corresponding to localization
length $\xi$). Thus we can define a length scale $\xi_c$ (as indicated
in the figure) at which there is a cross-over from the initial slope to the 
asymptotic slope. In the inset of Fig. \ref{ltl} we have shown the dependence 
of $\xi_c$ on $\eta$. Numerical fit shows that $\xi_c$ scales
as $1/\sqrt{\eta}$, as we expect  $\xi_c \rightarrow \infty$ with $\eta 
\rightarrow 0$. As one decreases $\eta$, the absolute value of initial slope increases
and that of the asymptotic one decreases. Simultaneously, the cross-over length
$\xi_c$ increases. In the $\eta \rightarrow 0$ limit both initial as well
as asymptotic slopes become identical.

\begin{figure}
\protect\centerline{\epsfxsize=6in \epsfbox{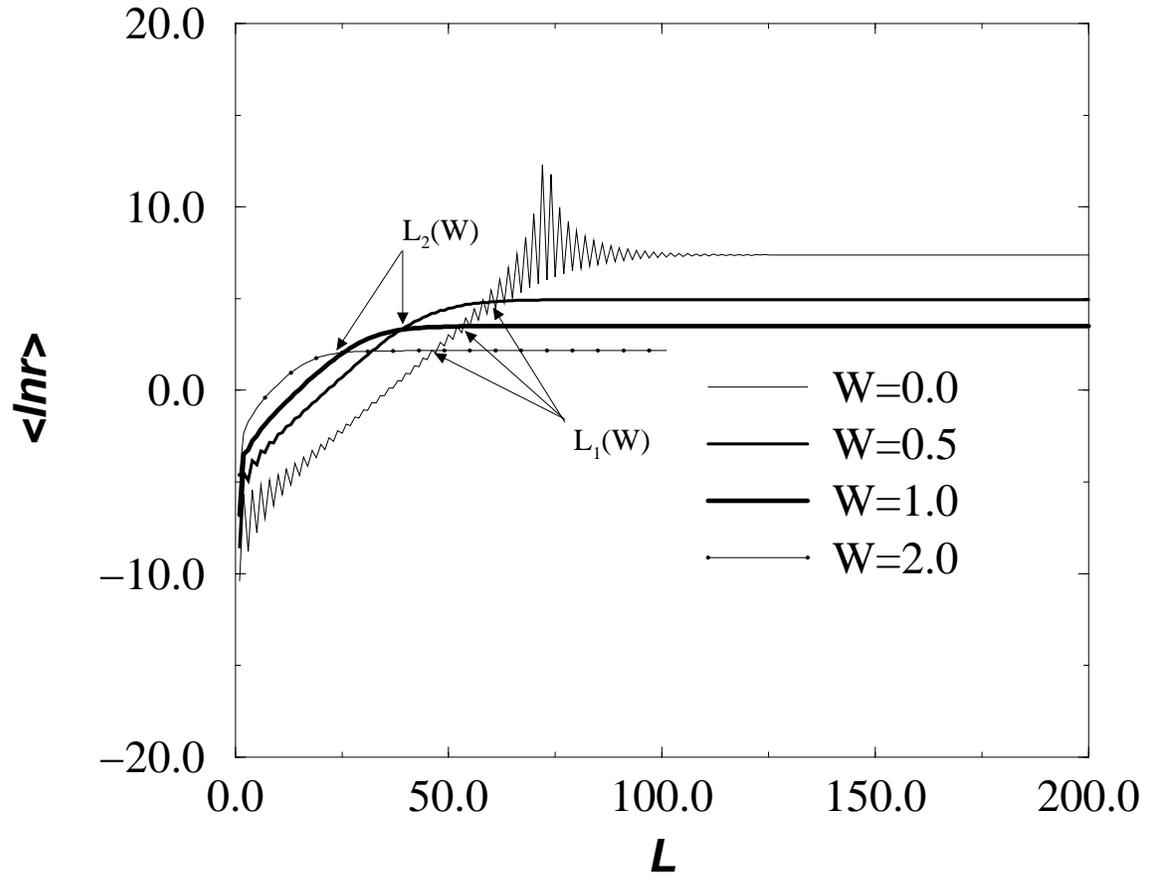}}
\caption{Variation of $\avg{lnr}$ with $L$ for values of $W$ indicated in 
the figure. The two length scales $L_1(W)$ and $L_2(W)$ associated with the 
reflectance are shown with arrows.}
\label{lnrvsl}
\end{figure}

In Fig. \ref{lnrvsl} we plot $\avg{lnr}$ as a function of the length $L$ 
for a fixed value of amplification strength $\eta=0.1$ and for various 
values of the disorder strength $W$ as indicated in the figure. In the 
absence of disorder ($W=0$) as one varies length, initially the 
reflectance increases to a very large value through large oscillations 
and after exhibiting a maximum again through oscillations, it eventually 
saturates to a finite (large) value. In the presence of disorder one can 
readily notice that initially $\avg{lnr}$ increases and has a magnitude 
larger than that for $W=0$ case and asymptotically beyond a disorder 
dependent length scale $L_1(W)$, it saturates to a value which is smaller 
than that for a $W=0$ case. The magnitude of the saturation value of 
$\avg{lnr}$ decreases as one increases the disorder as a result of 
localization induced by combined effect of disorder and amplification. 
Below the length scale $L_1(W)$ we can identify another disorder 
dependent length scale $L_2(W)$. Above $L_2$ (but smaller than $L_1$) 
further increase in disorder suppresses the reflectance whereas 
below $L_2$ it enhances the reflectance. The length scale $L_2$ 
being much smaller than the localization length $l$ for the passive 
medium, increase in disorder causes multiple reflections in a sample of 
size smaller than $L_2$ and consequently due to the increase in delay 
time we get enhanced back reflection. Beyond $L_2$ due to disorder 
induced localization delay time decreases and as a consequence we obtain 
reduced reflectance.

\begin{figure}
\protect\centerline{\epsfxsize=6in \epsfbox{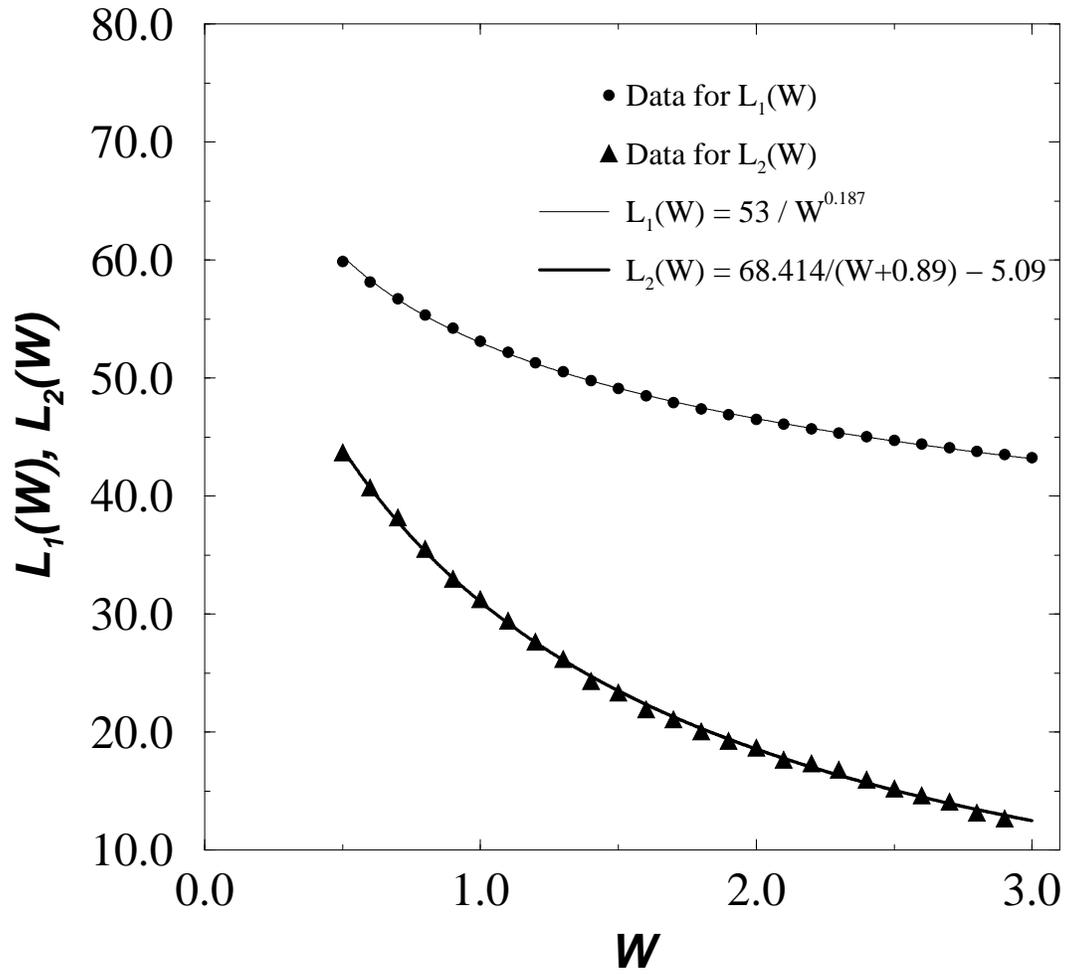}}
\caption{The variation of $L_1$ and $L_2$ with disorder strength $W$ for a 
sample with fixed amplification strength $\eta=0.1$.}  
\label{l1l2vsW}
\end{figure}

\begin{figure}
\protect\centerline{\epsfxsize=6in \epsfbox{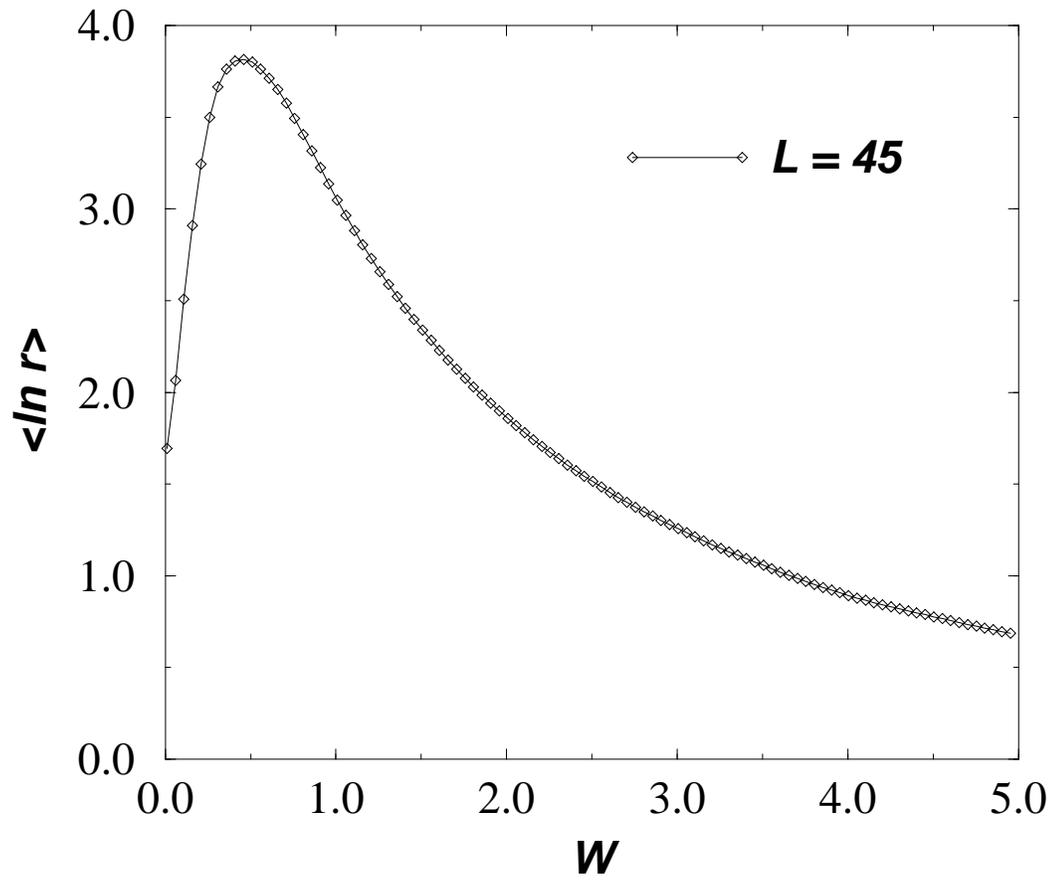}}
\caption{The variation of $\avg{lnr}$ with disorder $W$ for a sample of length
$L=45$ and $\eta=0.1$.}
\label{lnrvsW}
\end{figure}

The dependence of $L_1$ and $L_2$ on $W$ for a fixed value of $\eta=0.1$ 
is shown in Fig. \ref{l1l2vsW}. Both these length scales decrease as we 
increase $W$. The magnitude of $L_2$ is closer to the value of $L_c$ for 
a given disorder. In calculating $L_1$ we encounter error bars as the 
reflectance of a perfect system ($W=0$) exhibits oscillations. Curve 
fitting for larger values of disorder indicates  that $L_1$ and $L_2$ 
decreases as $1/W^{0.187}$ and $1/W$ respectively. From the existence of 
$L_1$ and $L_2$ one can readily infer that, if we have a sample of fixed  
length $L$ and amplification $\eta$, then as we increase disorder $W$ 
first due to multiple reflection (sample size $L$ being less than $L_2$) 
reflectance will increase. When the disorder strength becomes large such 
that $L_2 < L$ disorder induced localization will reduce the 
reflectance. This is shown in Fig. \ref{lnrvsW} where we plot $\avg{lnr}$ 
versus $W$ for a sample of fixed length $L=45$ and amplification $\eta=0.1$. 

\begin{figure}
\protect\centerline{\epsfxsize=6in \epsfbox{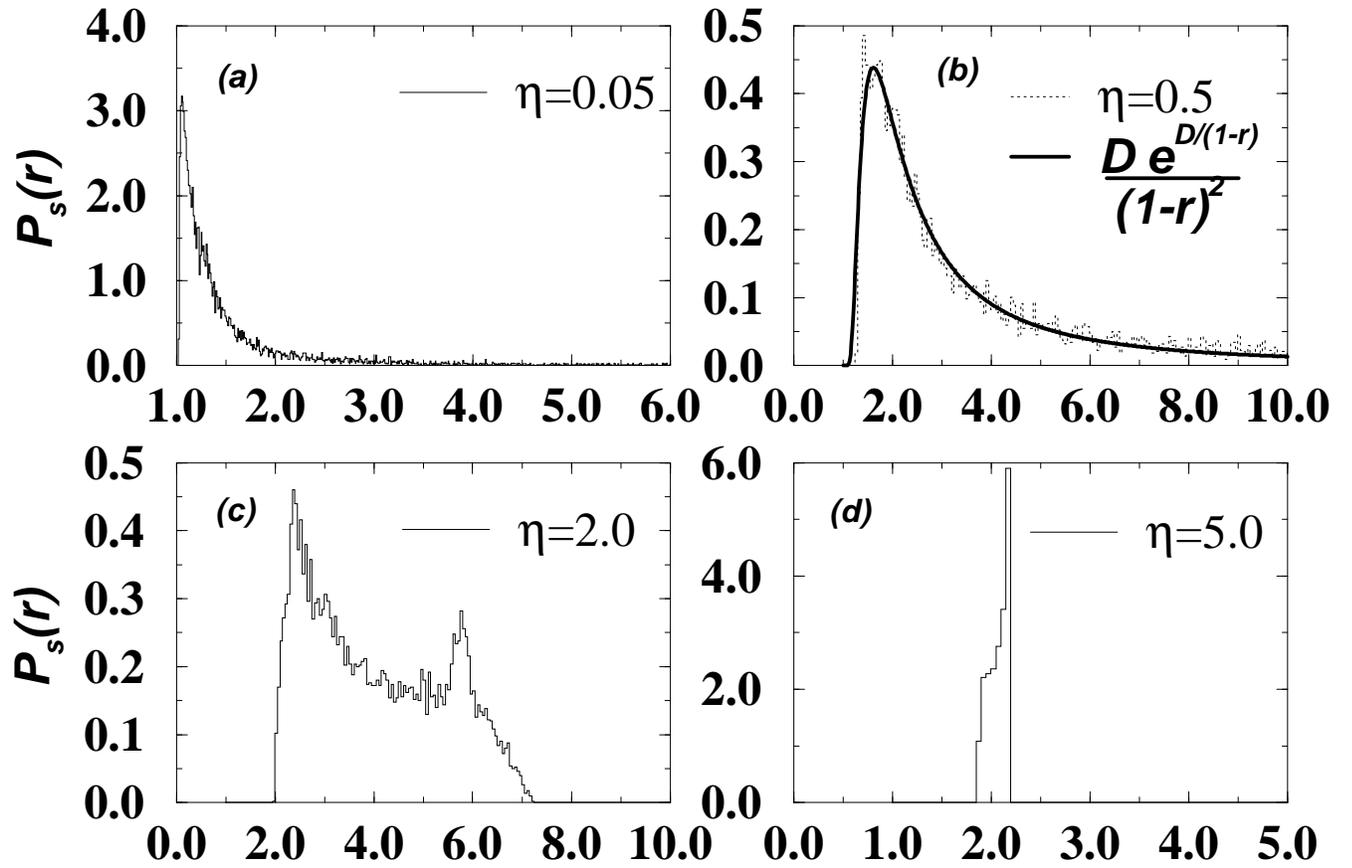}}
\caption{Stationary distribution of reflection coefficient $P_s(r)$ for $W=5.0$ and 
various values of $\eta$. The numerical fit shown in Fig. \ref{psr}(b)
with a thick line has $D=1.235$}
\label{psr}
\end{figure}

In Fig. \ref{psr} we have plotted the stationary distribution $P_s(r)$ of
reflection coefficient $r$ for different values of $\eta$ (as shown in the
figure) and a fixed value of $W=5.0$. To obtain stationary distribution we have
considered sample sizes much larger than the localization length $\xi$ such that
any increment in the length does not change the distribution. For small values
of $\eta=0.05$ the stationary distribution $P_s(r)$ has a single peak around
$r=r_{max}=1$. The peak ($r_{max}$) shifts to higher side as we increase $\eta$
(Fig. \ref{psr}(b)). The behavior of $P_s(r)$ for small $\eta$ is in qualitative
agreement with Eqn. (\ref{prad}). As we increase $\eta$ further $P_s(r)$ 
exhibit a double peaked structure (Fig. \ref{psr}(c)). At first the second
peak appears at higher value of $r$ at the expense of the distribution at 
the tail. As we increase $\eta$ the second peak becomes more prominent and 
shifts towards left, where as the height of the first peak decreases. The
distribution at the tail has a negligible weight (see Fig. \ref{psr}(c)). At
still higher values of $\eta$, the second peak approaches $r \approx 1$ whereas 
the first peak disappears. The now-obtained single peak distribution $P_s(r)$ in
the large $\eta$ limit tends to $\delta(r-1)$. In this limit the amplifying medium 
acts as a reflector and the disorder plays a sub-dominant role. The occurrence of
the double peak structure along with $P_s(r) \rightarrow \delta(r-1)$ in the large
$\eta$ limit cannot be explained even qualitatively from Eqn. (\ref{prad}).
This is due to the failure of RPA in this regime.

\begin{figure}
\protect\centerline{\epsfxsize=6in \epsfbox{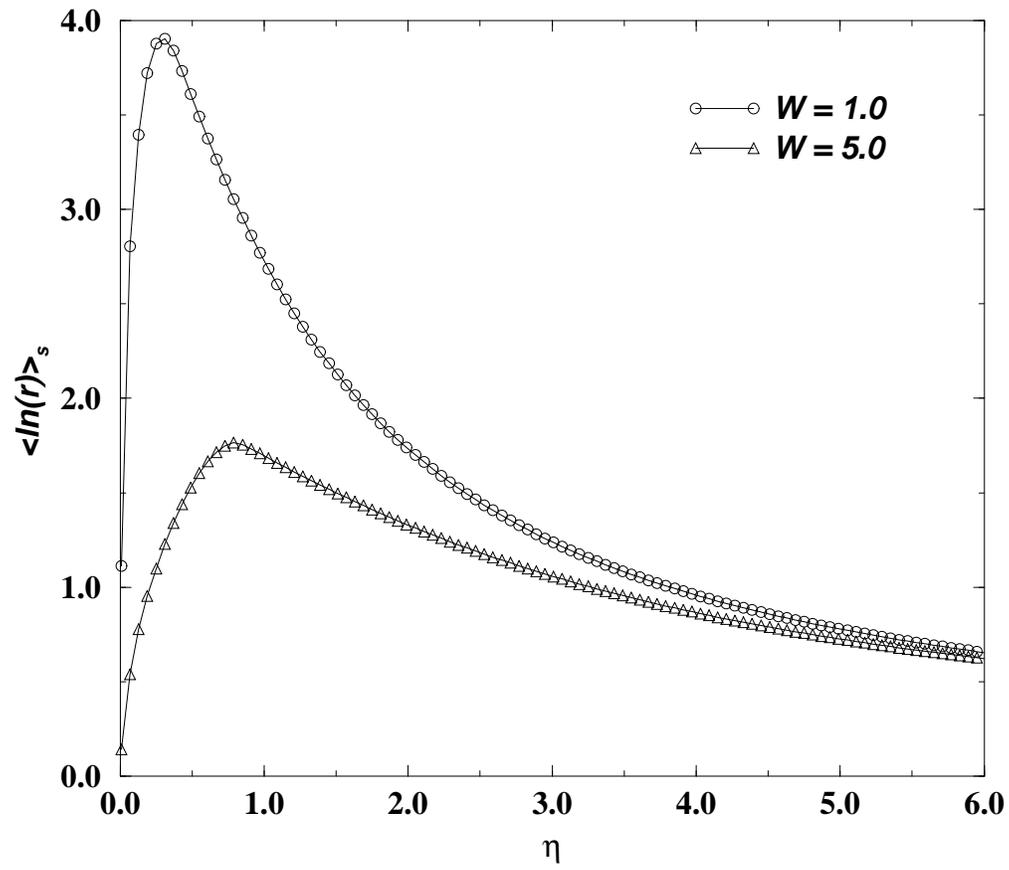}}
\caption{Variation of $\avg{ln(r)}_s$ with $\eta$ for 
two values $W$ as indicated in the figure.}
\label{reta}
\end{figure}

In Fig. \ref{reta} we have plotted $\avg{lnr}_s$, obtained from $P_s(r)$, as
a function of $\eta$ for $W=1.0$ and $W=5.0$ as indicated in the figure. As we
increase the strength of $\eta$, $\avg{lnr}_s$ first increases and after
exhibiting a maximum at $\eta=\eta_{max}$, $\avg{lnr}_s$ decreases 
monotonically. The numerical value of $\eta_{max}$ depends on the material
parameters. For the value $\eta > \eta_{max}$, the amplifying medium acts
dominantly as a reflector. It should be noted that the double peak structure in
$P_s(r)$ appears for values of $\eta$ close to $\eta_{max}$. For a given 
amount of disorder and for $\eta < \eta_{max}$ the increase in reflectance 
(beyond unity) as function of $\eta$ is due to the presence of disorder along with 
amplification. The randomness leads to multiple reflections and as a consequence
particles spend large amount of time in the sample before getting coherently
reflected. This enhances the total reflection and the peak value of $P_s(r)$
shifts to higher values of $r$. Beyond $\eta_{max}$, the amplifying medium plays a
dominant role as a reflector, leading to decrease in $\avg{lnr}_s$. Physics of
the double peak and overall shape of $P_s(r)$ shows similarity with the 
stationary distribution obtained in the case of absorption (for details we
refer to Ref. \cite{abhi}).

\begin{figure}
\protect\centerline{\epsfxsize=6in \epsfbox{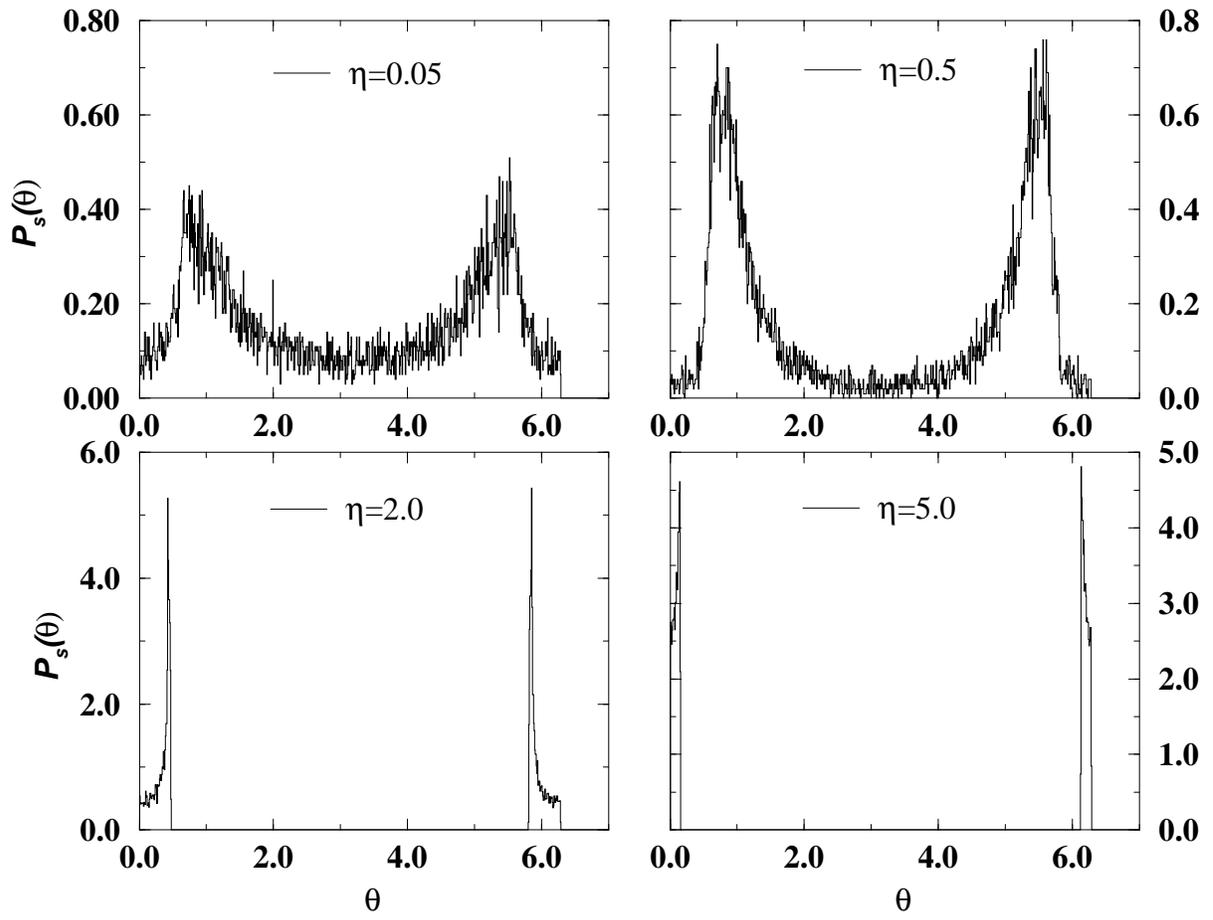}}
\caption{Stationary distribution of the phase of reflection amplitude
$P_s(\theta)$ versus $\theta$.}
\label{psph}
\end{figure}

Fig. \ref{psph} shows the stationary distribution $P_s(\theta)$
of the phase $\theta$ of the complex reflection amplitude for different values
of $\eta$. For the sake of convenience we have used the same parameters as in
Fig. \ref{psr}. It is clear from this figure that as we increase $\eta$ phase
fluctuations are suppressed. Double peak distribution is obtained even for
small $\eta$ (Fig. \ref{psph}(a)). With increasing $\eta$ the peaks become
prominent and they move apart. In the large $\eta$ limit $P_s(\theta)$ tends
to $\delta(\theta)$ and $\delta(\theta+2\pi)$. It should be noted that only in the 
limit $W<1$ and $\eta<1$ we obtain a uniform phase distribution over the range $0$ to
$2\pi$. It is this suppression of phase fluctuations in the presence of
amplification which leads to the breakdown of RPA. Hence the results based on RPA cannot 
explain the observed distribution of $P_s(r)$ at large $\eta$ (Fig. \ref{psr}(c,d)), 
even qualitatively. 
 
\section{CONCLUSIONS}

Our numerical study on the statistics of transmission coefficient in
random lasing medium indicates that in the asymptotic regime the
transmission coefficient is a non-self-averaging quantity,however, with a
well defined finite average value. We have shown that disorder suppresses
the transmittance at all length scales for a given fixed $\eta$. In some
parameter space transmittance initially increases with $\eta$ and falls
off exponentially to zero in the asymptotic regime. In this regime there
is a well defined cross-over length $L_c$ at which the transmittance is
maximum, and it decreases monotonically with $\eta$ as well as $W$. In the
parameter range where $\eta \ll 1$, in the presence of disorder the
average transmittance decreases monotonically and has a magnitude less
than unity. In this regime $L_c$ does not exist. However, one can still
define a new length scale $\xi_c$ which scales as $1/\sqrt{\eta}$. We have
also shown that there are two more length scales $L_1(W)$ and $L_2(W)$
associated with reflectance. For a system size upto $L_2(W)$ disorder
increases reflectance, for $L_2(W) < L < L_1(W)$ disorder suppresses the
reflectance. However, in this regime the reflectance is larger than that
for an ordered lasing medium. For $L > \L_1(W)$ disorder suppresses the
saturated value of $\avg{lnr}$ to a value much less than that for the case
of the ordered lasing medium. Our results clearly indicate that to obtain
an enhanced back reflection for a sample of fixed length $L$, the synergy
between the disorder and the amplification is necessary. The nature of the
stationary distribution of reflection coefficient $P_s(r)$ indicates that
earlier analytical studies fail, even qualitatively, to explain the
observed behavior in the large $\eta$ limit. In this limit, one can show
that amplification suppresses the phase fluctuations of complex reflection
amplitude and thus RPA is no longer valid. Our study clearly brings out
the dual role played by an amplifying medium, as an amplifier as well as a
reflector.


\end{document}